\documentclass[MSNbibl,seceqn,dvips]{arxstspdf}
\usepackage{dcolumn}
\usepackage{flushend}
\usepackage{stfloats}
\usepackage{graphicx}

\volume{27}
\issue{2}
\pubyear{2012}
\firstpage{294}
\lastpage{307}
\doi{10.1214/11-STS372}

\makeatletter
\newcommand{\eqref}[1]{(\ref{#1})}

\newcolumntype{d}[1]{D{.}{.}{#1}}
\newproclaim{example}{Example}[section]
\newproclaim{remark}{Remark}[section]
\newproclaim{thmm}{Theorem}[section]
\makeatother

\begin{document}
\begin{frontmatter}

\title{The Interval Property in Multiple Testing of Pairwise Differences}
\runtitle{Interval Property in Multiple Testing}

\begin{aug}
\author[a]{\fnms{Arthur} \snm{Cohen}\ead[label=e1]{artcohen@rci.rutgers.edu}}
\and
\author[b]{\fnms{Harold} \snm{Sackrowitz}\corref{}\ead[label=e2]{sackrowi@rci.rutgers.edu}}
\runauthor{A. Cohen and H. Sackrowitz}

\affiliation{Rutgers, The State University of New Jersey}

\address[a]{Arthur Cohen is Professor, Department of Statistics and
Biostatistics, Rutgers University, Piscataway, NJ 08854, USA \printead{e1}.}
\address[b]{Harold Sackrowitz is Professor, Department of Statistics and
Biostatistics, Rutgers University, Piscataway, NJ 08854, USA \printead{e2}.}
\end{aug}

%
\begin{abstract}
The usual step-down and step-up multiple testing procedures most often
lack an important intuitive, practical, and theoretical property called
the interval property. In short, the interval property is simply that
for an individual hypothesis, among the several to be tested, the
acceptance sections of relevant statistics are intervals. Lack of the
interval property is a serious shortcoming. This shortcoming is
demonstrated for testing various pairwise comparisons in multinomial
models, multivariate normal models and in nonparametric models.

Residual based stepwise multiple testing procedures that do have the
interval property are offered in all these cases.
\end{abstract}

%

\vspace*{-3pt}
\end{frontmatter}

\section{Introduction}\label{sec1}

Stepwise multiple testing procedures are valuable because they are
less conservative than standard single-step procedures which often rely
on Bonferroni critical values. In other words, they are more powerful
than their single-step counterparts. In constructing stepwise testing
procedures it is common to begin with tests for the individual
hypotheses that are known to have desirable properties. For example,
the tests may be UMPU, they may have invariance properties and are
likely to be admissible. Then a sequential component is added that
tells us which hypotheses to accept or reject at each step and when to
stop. We begin with the realization that all stepwise procedures induce
new tests on\vadjust{\goodbreak} the individual testing problems. Carrying out a stepwise
procedure in a multiple hypothesis testing problem is equivalent to
applying these induced tests separately to the individual hypotheses.
Thus, if the induced individual tests can be improved, then the entire
procedure is improved. Due to the sequential component, the nature of
these induced tests is typically complicated and overlooked.
Unfortunately they frequently do not retain all the desirable
properties that the original tests possessed.

In this paper we focus on an important type of practical property
(which in many models is also a necessary theoretical property) that we
call the interval property. This is a desirable property that the
original tests would typically have but that the stepwise induced tests
can easily lose. Informally the interval property is simply that the
resulting test has acceptance sections that are intervals.

To further clarify, suppose one is constructing a~test for a one-sided
hypothesis testing problem. In addition to asking for other properties
it is sensible to examine the acceptance and rejection regions. There
are often pairs of sample points, $\mathbf{X}$ and $\mathbf{X}^\ast$,
for which
there are compelling practical (and sometimes theoretical) reasons for
the following to be true. If the point $\mathbf{X}$ is in the rejection
region, then the point $\mathbf{X}^\ast$ should also be in the rejection\vadjust{\goodbreak}
region. The practical desirability of this property is usually due to
the fact that it is intuitively ``clear'' that $\mathbf{X}^\ast$ is a
stronger indication of the alternative than is $\mathbf{X}$. In the
case of
two-sided hypotheses there are often triples of points, $\mathbf{X},
\mathbf
{X}^\ast$ and $\mathbf{X}^{\ast\ast}$ (on the same line), such that if
both \textbf{X} and $\mathbf{X}^{\ast\ast}$ are in the acceptance region, then one
would also want $\mathbf{X}^\ast$ to be in the acceptance region if in fact
$\mathbf{X}^\ast$ was not the most indicative of the alternative of the
three points.

We illustrate this idea with an example that will be treated fully in
Section \ref{sec5.1}. Suppose one observes the data in Table~\ref{tab1}
based on the
three labeled independent treatments. One of the hypotheses of interest
is whether or not the distribution for Dose 1 is stochastically larger
than that for the placebo. If the method used decides in favor of
stochastic order based on observing Table~\ref{tab1}, then it should also
decide in favor of Dose 1 if Table \ref{tab2} is observed. Repeated use
of a
test procedure not having this property will ultimately lead to
conclusions that seem contradictory and would be difficult to justify.
The interval property is not only natural but is necessary for
admissibility. We will return to Tables \ref{tab1} and \ref{tab2} later
in Section \ref{sec5.1}.

\begin{table}
\caption{Health Status data at sample point $\mathbf{x}$}\label{tab1}
\begin{tabular*}{\columnwidth}{@{\extracolsep{\fill}}lcccc@{}}
\hline
& \textbf{Same} &\textbf{Improved}& \textbf{Cured} & \\
\hline
Placebo&15&226&\phantom{0}4&245\\
Dose 1&\phantom{0}4&226&15&245\\
Dose 2&\phantom{0}6&196&43&245 \\
\hline
\end{tabular*}
\end{table}

\begin{table}[b]
\caption{Health Status data at sample point $\mathbf{x}^{\bolds{\ast}}$}\label{tab2}
\begin{tabular*}{\columnwidth}{@{\extracolsep{\fill}}lcccc@{}}
\hline
& \textbf{Same} &\textbf{Improved}& \textbf{Cured} & \\
\hline
Placebo&16&226&\phantom{0}3&245\\
Dose 1&\phantom{0}3&226&16&245\\
Dose 2&\phantom{0}6&196&43&245 \\
\hline
\end{tabular*}
\end{table}

We study this idea in the most common of multiple testing situations,
that is, those where hypotheses under consideration involve collections
of pairwise differences. The most common of these are\break (i)~treatments
versus control problems, (ii) change point problems and (iii) problems
examining all pairwise differences. We will investigate these problems
in a~broad spectrum of models: univariate models involving means or
variances, multivariate models concerning mean vectors, ordinal data
models involving\vadjust{\goodbreak} equality of multinomial distributions and
nonparametric models involving equality of distributions.

Two popular types of multiple testing procedures for such problems are
a step-down procedure (to be defined later) and a step-up procedure. To
simplify the presentation we focus mainly on the step-down procedure as
analogous results can be obtained for the FDR controlling step-up
procedure of \citet{r3}. We will see that these step-down induced tests
often do not retain the interval property. In fact, among all the
models considered the usual step-down procedure maintains the interval
property only when testing treatments versus control in the one-sided
case. We will also show how to construct a step-down procedure that
does have the interval property. Furthermore, it should be clear from
the examples and from the way that the methods are used that this
phenomenon exists in a far greater variety of models.

The usual step-down procedure is given in \citet{r16}. For testing all
pairwise comparisons variations are offered in \citet{r15}, \citet{r19},
\citet{r18} and \citet{r20}. The lack of the interval property in a
one-way ANOVA model for testing all pairwise contrasts is shown in
\citet
{r9} (CSC) under a normal model. It has also been demonstrated for
rank tests in a one-way ANOVA model in \citet{r8} (CS).

Many multiple testing procedures are designed to control some error
rate such as the familywise error rate FWER (weak and strong), the
false discovery rate FDR and k-FWER (see \citep{r16}). Some researchers
also take a finite action decision theory problem approach with a
variety of loss functions (e.g., \citep{r13}). In these studies
procedures are evaluated and compared by their risk functions. The risk
function approach does not always necessitate the need to control a
particular type of error rate. \citet{r12} study expected values of
functions of numbers of Type I and Type II errors. In any particular
application one would typically have a sense of desirable criteria as
well as those portions of the parameter space that are most relevant.
To get a more complete understanding of the behavior of one's procedure
we recommend that, if feasible, error control and risk function
properties should be examined.

In this paper we specify procedures that have the interval property
for a much wider class of both univariate and multivariate models. For
exponential family models, where individual test statistics are
dependent, each individual test induced by usual step-down and step-up
procedures has been shown to be inadmissible with respect to the
classical hypothesis testing \mbox{0--1} loss. See Cohen and Sackrowitz
(\citeyear{r5,r6,r7}) and CSC
(2010) cited above.\break Those proofs are based on
results of \citet{r17} that, in effect, say that the interval property
is equivalent to admissibility. One implication of this is that no
Bayesian approach would lead to a procedure that lacks the interval
property. Thus no prior distribution can be used to explain a~lack of
the interval property.

Lack of the interval property not only means that, in exponential
family models, procedures exist with both better size and power for
\textit{every} individual hypothesis, but it may also lead to very
counterintuitive results. It is hard to believe a client would be happy
with a procedure that could yield a reject of a null hypothesis in one
instance and then yield an accept of the same hypothesis in another
instance when the evidence and intuition is more intuitively compelling
in the latter case.

The methodology we present leads to procedures that are admissible.
Furthermore, their operating characteristics often compare favorably
with the\break usual step-down procedures. This behavior can be seen from the
simulations presented in \citet{r11} (CSX). In that same paper a family
of residual based procedures were defined. The step-down procedures
having the interval property that will be presented in this paper stem
from those procedures. They are exhibited in special cases in CSC
(2010) and CS (\citeyear{r8}).

In the models considered here, the Residual based Step-Down procedures,
labeled RSD, exhibit two important characteristics. It begins with the
set $S = \{1, 2, \ldots, k\}$ where each integer is associated with a~population. Next, based on all the data, $S$ is partitioned into a
collection of disjoint sets through a sequential process. Finally,
hypothesis $H_{ij}$ (that population \textit{i} is equal to population
\textit{j}) is accepted if and only if both $i$ and $j$ are in the same
set of the final partition. Second, the partitioning process is based
on the pooling of various samples (depending on the particular model at
hand) at each stage. The final partition of the set is reached through
a sequence of partitions that become finer at each step

There are some noteworthy differences between step-up or step-down and
RSD. Depending on the collection of hypotheses being tested, there will\vadjust{\goodbreak}
be correlation between many of the test statistics being used. Neither
step-up nor step-down allows for this in the construction of the test
statistic itself. Thus those test statistics will be the same
regardless of the correlation structure. The RSD methodology yields
statistics that are determined by the correlation structure.
Furthermore, the RSD test statistics change at each step depending on
the actions taken at the previous step.

Unfortunately, insight as to why the interval property will ensue in
some cases but not others is still wanting. The crucial element seems
to be the way the test statistics and stopping rules mesh and this must
be checked mathematically.

We point out that many of the step-down procedures discussed here are
symmetric in the sense that whatever is true for any one hypothesis to
be tested is also true for the other hypotheses to be tested. So
although the lack of the interval property is shown for one particular
testing problem, it is true for all individual problems. This takes on
added significance for exponential family models. It means that every
individual test is inadmissible. When the number of hypotheses is
large, the number of opportunities for inconsistent decisions also gets
to be large. For risk functions that would sum mistakes, such as the
classification risk (\citep{r13}), this could amount to considerable error.

Lastly, we mention the issue of critical values. The shortcoming of RSD
and to some extent all stepwise procedures is in determining sharp
critical values. This is particularly true in the face of dependence
which is exactly the situations in which usual stepwise procedures tend
to lack the interval property. With knowledge (based on practicality)
of relevant criteria and relevant portions of the parameter space as
focus, one can search for appropriate critical values using
simulations. A good first simulation for RSD is to use the critical
values suggested in the work of \citet{r2} and modify them if necessary.
The standard step-up and step-down procedures do not take dependency
into account in choosing a level and can also benefit by using
simulation to modify their critical values. As examples, two
simulations are given for a simple model in Section \ref{sec6.3}. There we
compare RSD and step-up in a treatments versus control setting.

In the next section we give models and definitions. Several models, for
which the results of the paper hold, are listed. These include normal
models, multinomial models, and arbitrary continuous distribution\vadjust{\goodbreak}
models treated nonparametrically. Section~\ref{sec3} discusses counterexamples
to the interval property. In Section~\ref{sec4} we introduce a
step-down method,
called RSD, that leads to procedures that do have the interval
property. Sections \ref{sec5}, \ref{sec6} and \ref{sec7} contain
results for multinomial models,
multivariate normal models and nonparametric models, respectively.

\section{Models and Definitions}\label{sec2}

Let $\pi_i, i=1, \ldots, k$, be $k$ independent populations. Data from
population $\pi_i$ is denoted by a $q \times1$ vector~$\mathbf{X}_i$
and $\mathbf{X}$ represents $(\mathbf{X}_1^\prime, \ldots,\mathbf
{X}_k^\prime)^\prime$.

Hypotheses of interest, for particular (\textit{i}, \textit{j})
combinations, are denoted by $H_{ij} \dvtx \pi_i= \pi_j$ versus
$K_{ij}\dvtx \pi
_i \neq\pi_j$ or $K_{ij}\dvtx \pi_i < \pi_j$. The latter one-sided case
can be interpreted as the difference in two scalar parameters in case
$\pi_i$ is characterized by a single parameter or $<$ can be
interpreted as $\pi_j$ is stochastically larger than $\pi_i$ in case
$\pi_i$ are multinomial distributions or other distributions not
necessarily characterized by parameters. We consider situations where
there are at least two connected hypotheses among those to be tested,
that is, an $H_{ij}, H_{jm}$ or an $H_{ij}, H_{im}$. We study the
following three problems in the domain of pairwise differences:

\begin{enumerate}
\item All pairwise differences. Here $H_{ij} \dvtx \pi_i = \pi_j$ versus
$ K_{ij} \dvtx \pi_i \neq\pi_j,$ all $i < j, i,j = 1, \ldots, k.$

\item Change point. $H_{i(i+1)} \dvtx \pi_i = \pi_{i+1}$ versus
$K_{i(i+1)} \dvtx \allowbreak \pi_i < \pi_{i+1}, i=1, \ldots, k-1,$ where $<$ can
mean sto\-chastically less than or if $\pi_i$ is characterized by a
parameter it simply means that the parameter for population $i$ is less
than the parameter for population $i+1.$ Two-sided alternatives can
also be considered.

\item Treatments versus control. $H_{ik} \dvtx \pi_i = \pi_k$ versus
$K_{ij} \dvtx  \pi_i \neq\pi_k, i= 1, \ldots, k-1$.
\end{enumerate}

Problems 1, 2 and 3 will be studied for the following probability models:

\begin{enumerate}

\item$\pi_i$ are independent multinomial distributions. For problem 2
assume $\pi_1 \leq\pi_2 \leq\cdots\leq\pi_k$ so that the
alternative hypotheses are strict stochastic order.

\item$\pi_i$ are independent $p$-variate normal distributions with
unknown mean vectors $\bolds{\mu}_i$ and known covariance matrix
$\Sigma$.

\item Assume $\pi_i$ has c.d.f. $F_i$ with $F_i$ continuous. For problem
2 assume $F_1 \leq\cdots\leq F_k$ so alternatives are strict
stochastic order.
\end{enumerate}

The intuitive description of the interval property given in Section \ref{sec1}
will be given a formal interpretation on a case by case basis as
follows. In each specific model, when $H_{ij}$ is being tested, a\vadjust{\goodbreak}
vector~$\mathbf{g}_{ij}$ will be identified based on compelling
practical (and/\allowbreak or theoretical) considerations so that a nonrandomized
test $\varphi_{ij}(\mathbf{x})$ will be said to have the interval
property (relative to the identified $\mathbf{g}_{ij}$) if $\varphi
_{ij}(\mathbf{x} + a\mathbf{g}_{ij})$

\begin{longlist}[(ii)]
\item[(i)] is nondecreasing as a function of $a$ in the one-sided case,

\item[(ii)] has a convex acceptance region in $a$ in the two-sided case.
\end{longlist}

These practical considerations turn out to involve only the data coming
from the populations $\pi_{i}$ and $\pi_{j}$ as they are independent of
all the other populations. Thus $ \mathbf{g}_{ij}$ will be seen to have
entries of 0 for all coordinates that do not correspond to data from
$\pi_{i}$ or $\pi_{j}$. Let $\widehat{\mathbf{g}}_{ij}$ be the $2q
\times1$ vector consisting of the elements of $ \mathbf{g}_{ij}$ that
pertain to $\pi_{i}$ and $\pi_{j}$.

Now let $\widehat{T}_{ij}(\mathbf{x}_{i},\mathbf{x}_{j})$ be the
two-population test sta\-tistic for testing $H_{ij}$ that, when only
$(\mathbf{x}_{i},\mathbf{x}_{j})$ are observed, is the basis of the
usual step-down procedure. When all of $\mathbf{x}$ is observed we
define $T_{ij}(\mathbf{x}) = \widehat{T}_{ij}(\mathbf{x}_{i},\mathbf
{x}_{j})$. That is, $T_{ij}$ is a function that depends on $\mathbf{x}$
only through $(\mathbf{x}_{i},\mathbf{x}_{j})$.

Also let $\widehat{\psi}_{ij}((\mathbf{x}_{i},\mathbf{x}_{j}))$ be the
nonrandomized test function which utilizes $\widehat{T}_{ij} (\mathbf
{x}_{i},\mathbf{x}_{j})$. That is, for a one-sided test $\widehat{\psi
}_{ij}(\mathbf{x}_{i},\mathbf{x}_{j})= 1$ if $\widehat{T}_{ij}((\mathbf
{x}_{i},\mathbf{x}_{j})) > C$ and $\widehat{\psi}_{ij}(\mathbf
{x}_{i},\mathbf{x}_{j})= 0$ otherwise. For a two-sided test $\widehat
{\psi}_{ij}(\mathbf{x}_{i},\allowbreak\mathbf{x}_{j})= 1$ if $\widehat
{T}_{ij}(\mathbf{x}_{i},\mathbf{x}_{j}) < C_L$ or $\widehat
{T}_{ij}(\mathbf{x}_{i},\mathbf{x}_{j}) > C_U$. Otherwise $\widehat{\psi
}_{ij}(\mathbf{x}_{i},\mathbf{x}_{j})= 0$.

In the vast majority of multiple testing problems the same two-sample
test statistic is used for every~$H_{ij}$. To simplify notation we will
use this setting. Extension to the general case would follow easily.
Thus, when clear, we suppress subscript notation for two-sample
functions as follows:
\begin{eqnarray*}
\widehat{\mathbf{g}}_{ij} &=& \widehat{\mathbf{g}},\quad \widehat{T}_{ij}
(\mathbf{x}_{i},\mathbf{x}_{j}) = \widehat{T} (\mathbf{x}_{i},\mathbf
{x}_{j}) \quad\mbox{and}\\
\widehat{\psi}_{ij}(\mathbf
{x}_{i},\mathbf{x}_{j}) &=& \widehat{\psi}(\mathbf{x}_{i},\mathbf
{x}_{j}), \quad\mbox{all } i < j.
\end{eqnarray*}

We will say $\widehat{\psi}(\mathbf{x}_{i},\mathbf{x}_{j})$ has the
interval property relative to $\widehat{\mathbf{g}}$ in the two-sample
problem if $\widehat{\psi}_{ij}((\mathbf{x}_{i}', \mathbf{x}_{j}')' +
a\widehat{\mathbf{g}})$ satisfies (i) and (ii) above.

At this point we describe the usual step-down procedure for multiple
testing of a collection of hypotheses $H_{ij}$ based on statistics
$\widehat{T} (\mathbf{x}_i, \mathbf{x}_j)$. See, for example, \citet
{r11}. We describe the procedure for one-sided alternatives. For
two-sided alternatives sometimes statistics are absolute values or
upper and lower critical values are used. For one-sided alternatives
let $K$ be the number of hypotheses to be tested and let $0 \leq C_1 <
C_2 < \cdots< C_K$ be critical values. Define the collection of pairs
$Q= \{(i,j) \dvtx H_{ij}$ is to be tested$\}$.

\begin{longlist}[Step 1:]
\item[Step 1:] Let $\widehat{T}_{i_1,j_1} = \max_{(i,j)\in
Q} \widehat{T}(\mathbf{x}_i,\mathbf{x}_j).$ If $\widehat
{T}_{i_{1},j_{1}} \leq C_K,$ accept all hypotheses and stop.

If $\widehat{T}_{i_{1},j_{1}} > C_K,$ reject $H_{i_{1},j_{1}}$ and go
to step 2.

\item[Step 2:] Consider $\widehat{T}_{i_{2},j_{2}}\,{=}\,\max
_{(i,j)\in Q \setminus( i_{1}, j_{1})} \widehat{T}(\mathbf
{x}_i, \mathbf{x}_j)$. If $\widehat{T}_{i_{2},j_{2}} \leq C_{K-1},$ accept
all remaining hypotheses. If $\widehat{T}_{i_{2},j_{2}} > C_{K-1},$
reject $H_{i_{2},j_{2}}$ and go to step 3.

\item[Step $m$:] Consider
\[
\widehat{T}_{i_{m},j_{m}} = \max
_{(i,j)\in Q \setminus\{( i_{1}, j_{1}) \cdots(i_{m-1},
j_{m-1})\}} \widehat{T}(\mathbf{x}_i,\mathbf{x}_j).
\]

If $\widehat{T}_{i_{m},j_{m}} \leq C_{K-(m-1)},$ accept all remaining
hypotheses.

If $\widehat{T}_{i_{m},j_{m}} > C_{K-(m-1)},$ reject $H_{i_{m},j_{m}}$
and go to step~$(m+1)$.
\end{longlist}

We remark that the RSD methods presented are also based on the function
$\widehat{T}$. However, the arguments used are not $(\mathbf
{x}_{i},\mathbf{x}_{j})$.

\section{Prototype Counterexamples to the Interval Property}\label{sec3}

In this section we describe the fundamentals of searching for points at
which step-down procedures might violate the interval property. The
idea is to capitalize on a consequence of the sequential process as
follows. Suppose, when $\mathbf{x}$ is observed, the step-down
procedure rejects $H_{ij}$ based on the value of $T_{ij}(\mathbf{x})$
but does not do so until stage $m > 1$. Further suppose that when
$\mathbf{x}^{*}$ is observed there is even more evidence to reject
$H_{ij}$ based on $T_{ij}(\mathbf{x}^{*})$. The difficulty is that the
stopping rule may prevent the procedure from even reaching stage m when
$\mathbf{x}^{*}$ is observed.

\begin{figure}

\includegraphics{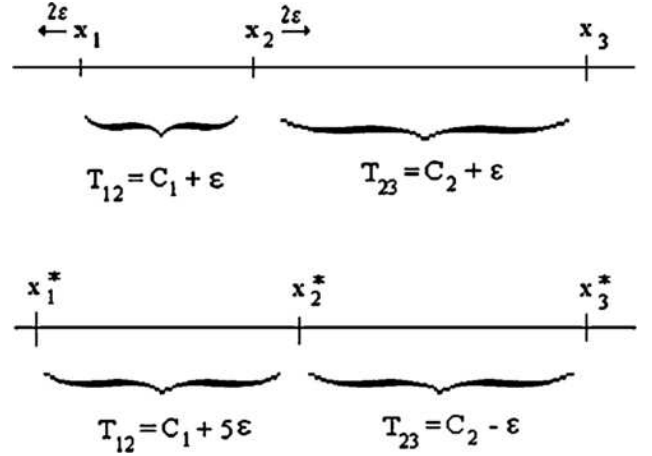}

\caption{Violation of interval property for one-sided change point
problem.}\label{fig1}
\end{figure}

\begin{figure}[b]

\includegraphics{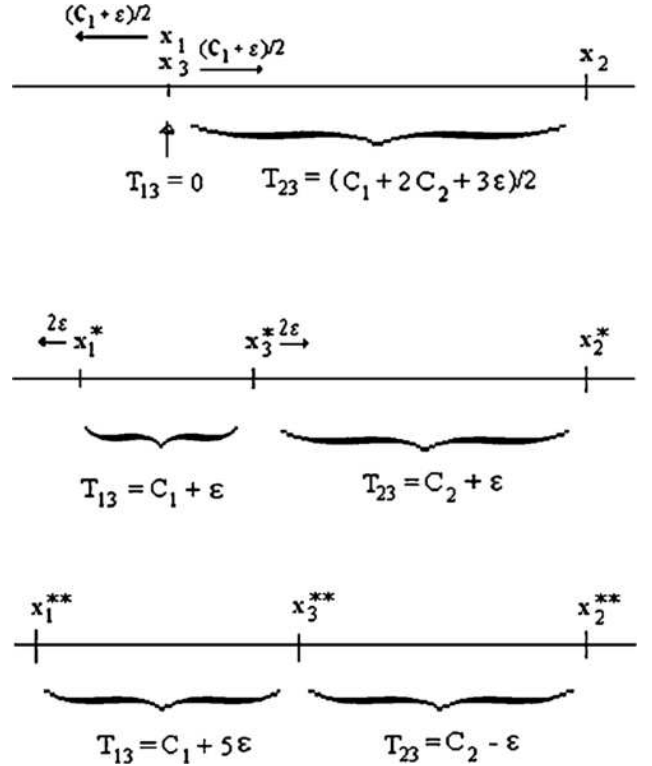}

\caption{Violation of interval property for two-sided treatments versus
control problem.}\label{fig2}
\end{figure}

To demonstrate we will consider some multiple testing situations using
only three populations $ \pi_{1}$, $\pi_{2}$, $\pi_{3}$. All the
fundamentals can be seen in the case that all $\mathbf{x}_{i}$ are
one-dimensional and $T_{ij} = \mathbf{x}_{j} - \mathbf{x}_{i}$ in the
one-sided case and $T_{ij} = |\mathbf{x}_{i} - \mathbf{x}_{j}|$ in the
two-sided case. Figures~\ref{fig1} and \ref{fig2} give an intuitive
sense of the sort of
behavior that one seeks for a~violation of the interval property. To
extend these ideas to more general situations we use the figures to
determine the desired relative positions (with distances measured by
the value of the test statistic) of sample points as one moves along
the sequence of points~$\mathbf{x}, \mathbf{x}^{*} \mbox{ and } \mathbf{x}^{**}$.

%

%

Figure \ref{fig1} is appropriate when the (change point) hypotheses to
be tested
are $H_{12} \dvtx \pi_1= \pi_2$ versus $K_{12}\dvtx\allowbreak \pi_1 < \pi_2$
and $H_{23}
\dvtx \pi_2= \pi_3$ versus\vadjust{\goodbreak} $K_{23}\dvtx \pi_2 < \pi_3$. Suppose
$\mathbf
{g}_{12} = (-1, 1, 0)$. When $\mathbf{x}$ is observed $H_{23}$ is rejected
at stage 1 and then $H_{12}$ is rejected at stage~2. When $\mathbf
{x}^{*} = \mathbf{x} + (2 \varepsilon) \mathbf{g}_{12}$ is observed
$H_{23}$ is now accepted at stage 1, causing the procedure to stop.
Thus $H_{12}$ is now accepted despite an increase in evidence against it.

Figure \ref{fig2} is appropriate when the (treatments versus control)
hypotheses to be tested are $H_{13} \dvtx \pi_1= \pi_3$ versus
$K_{13}\dvtx \pi
_1 \neq\pi_3$ and $H_{23} \dvtx \pi_2= \pi_3$ versus $K_{23}\dvtx\allowbreak \pi
_2 \neq
\pi_3$. Here $\pi_3$ is the control and $\mathbf{g}_{13} = (-1, 0, 1)$.
When $\mathbf{x}$ is observed $H_{23}$ is rejected at stage 1 and then
$H_{13}$ is accepted at stage 2. When $\mathbf{x}^{*} = \mathbf{x} +
((C_{1} + \varepsilon)/2) \mathbf{g}_{13}$ is observed $H_{23}$ is
rejected at stage 1 and then $H_{13}$ is rejected at stage 2. Finally,
when $\mathbf{x}^{**} = \mathbf{x}^{*} + ( 2 \varepsilon)) \mathbf
{g}_{13}$ is observed both hypotheses are accepted. In the sample space
as we go from $\mathbf{x}$ to $\mathbf{x}^{*}$ to $\mathbf{x}^{**}$ the evidence
against $H_{13}$ continues to mount. Yet the step-down procedure's
decisons are to accept, reject and then accept again on this sequence
of points.

Figure \ref{fig2} is also appropriate when testing all pairwise comparisons
provided $C_{1} + 2 C_{2} > 2 C_{3}$.

\section{RSD Features and First Properties}\label{sec4}

In this section we describe some specifics of the step-down procedures
we will present that do have the interval property. As previously
mentioned, decisions are, in effect, based on a final partition of the
set $S = \{1, 2, \ldots, k\}$ that is reached through a~sequence of
data based partitions that become finer at each step. Each integer is
associated with a population. Suppose the hypothesis $H_{ij} \dvtx \pi_i=
\pi_j$ is under consideration. Then $H_{ij}$ is rejected if and only if
$i$ and $j$ are in different sets of the final partition of~$S$. The
precise rules for the partitioning depend on the model and the data.
Illustrative examples of the process will be given at the end of this
section. However, certain principles are common to all
models.\looseness=1

At the first step the process either stops and $S$ itself is the final
partition (in this case no hypothesis can be rejected) or $S$ is
divided into two sets. At any future step the process either stops or
one of the sets in the current partition is divided into two nonempty
sets. The types of allowable sets in the partition process are often
restricted by the particular model being considered. For the process to
begin we must determine three (model-driven) classes of sets, $\Omega,
\Omega_{1}$ and $\Omega_{2}$. At any step only sets that lie in~$\Omega
$ are eligible to be split. Of course $\Omega$ must contain at least
two integers. One way the process will be stopped is if the current
partition contains no such sets. Further, if a set $B \in\Omega$ is to
be divided into~$A$ and $B\setminus A$ we require $A \in\Omega_{1}$
and $B\setminus A \in\Omega_{2}$. It is often the case that $ \Omega
_{1} = \Omega_{2}$. Whenever a set, say $B = \{i_{1}, \ldots, i_{m}\}
$, is under consideration to be split into two parts the decision is
based on some metric $H(A,B\setminus A;\mathbf{x})$ of set dispersion. Here
$H$ is defined only for $A \subset B$ with $A$ and\vadjust{\goodbreak} $B\setminus A$
both nonempty. For any set of integers, $A$, define
%
\begin{eqnarray}\label{eq4.1}
n(A)& =& \mbox{number of integers in } A\quad \mbox{and}
\nonumber
\\[-8pt]
\\[-8pt]
\nonumber
Y(A;\mathbf{x})& =& \sum_{j \in\mathrm{A}} x_{j}.
\end{eqnarray}

Due to the pairwise nature of each $H_{ij}$ the functions
$H(A,B\setminus A;\mathbf{x})$ used in the various multiple testing
problems will be chosen to depend only on the functions $n(\cdot)$ and
$Y(\cdot;\mathbf{x})$. Next let, for any \mbox{$B \subset\Omega$},
\[
D(B;\mathbf{x}) = \max_{A\subset B, A\subset\Omega
_{1}, B\setminus A \subset\Omega_{2}} H(A,B\setminus A;\mathbf{x})
\]
and let the max be attained for the set $A_{B}$. That is, $ D(B;\mathbf
{x}) = H(A_{B},B\setminus A_{B};\mathbf{x})$. If the set $B$ is ever to be
divided, it will be split into $A_{B}$ and $B\setminus A_{B}$. The
dependence of $A_{B}$ on $\mathbf{x}$ will usually be suppressed in the
notation.

Let $\{ C_{m}\}, m = 1,\ldots,k$ be an increasing set of critical
values. Suppose that for some sample point $\mathbf{x}$ stage \textit
{m} is
reached and the current partition entering stage \textit{m} is denoted
by $B_{1m}, \ldots, B_{mm}$. If
\[
\max(D(B_{1m};\mathbf{x}), \ldots, D(B_{mm};\mathbf{x}))
> C_{k+1-m}
\]
then split the set corresponding to the largest\break $D(B_{im}; \mathbf{x})$
and continue to the next stage. Otherwise stop.

This construction leads to the following two basic results.

\begin{thmm}\label{thm4.1} Suppose $H(A,B\setminus A;\mathbf{x}+a\mathbf{g})$ has the
following properties. It is
\begin{longlist}[(iii)]
\item[(i)] a nondecreasing function of a if $\{i\} \in A, \{j\} \in
B\setminus A$ or $\{j\} \in A, \{i\} \in B\setminus A;$

\item[(ii)] constant as a function of a if $\{i,j\} \subseteq A$ or $\{
i,j\} \subseteq B\setminus A;$

\item[(iii)] constant as a function of a if $\{i,j\} \cap B = \phi.$
\end{longlist}
If the final partition at the sample point $\mathbf{x}$ places~$i$ and $j$ in
different sets, then the final partition at $\mathbf{x}^{*} = \mathbf
{x} + a \mathbf{g}, a > 0$ will also place $i$ and $j$ in different sets.
\end{thmm}

\begin{pf}
Since the final partition at the point $\mathbf{x}$ placed \textit{i} and
\textit{j} in different sets, the partitioning process continued, at
least, until $i$ and $j$ were separated. Consider any stage in which
$i$ and $j$ have not yet been separated. In that partition let $B^{*}$
denote the set containing both $i$ and $j$. By assumptions (i)--(iii),
for any $B$ in that same partition we must have $H(A,B\setminus
A;\mathbf{x}+a\mathbf{g}) = H(A,B\setminus A;\mathbf{x})$ \textit{unless}
$B = B^{*}$ and $\{i\} \in A, \{j\} \in
B^{*}\setminus A$ or $\{j\} \in A, \{i\} \in B^{*}\setminus A$. In that
case if they are not equal, then [by (i)] we must have
$H(A,B^{*}\setminus A;\mathbf{x}+a\mathbf{g}) > H(A,B^{*}\setminus
A;\mathbf
{x})$. Thus \textit{i} and \textit{j} would become separated at the
point $\mathbf{x} + a \mathbf{g}$ at least as early as they were at the
point $\mathbf{x}$. The result now follows.
\end{pf}

\begin{thmm}\label{thm4.2} Suppose $H(A,B\setminus A;\mathbf{x}+a\mathbf{g})$ has the
following properties. It is
\begin{longlist}[(iii)]
\item[(i)] nonincreasing and then nondecreasing as\break a~function of a if $\{
i\} \in A, \{j\} \in B\setminus A$ or $\{j\} \in A,\allowbreak \{i\} \in
B\setminus A;$

\item[(ii)] constant as a function of a if $\{i,j\} \subseteq A$ or $\{
i,j\} \subseteq B\setminus A;$

\item[(iii)] constant as a function of a if $\{i,j\} \cap B = \phi.$
\end{longlist}

If the final partition at the sample point $\mathbf{x}$ places~$i$ and~$j$ in
the same set but the final partition at the sample point $\mathbf
{x}^{*} = \mathbf{x} + a_{1} \mathbf{g}, a_{1} > 0$ places~$i$ and~$j$ in
different sets, then the final partition at $\mathbf{x}^{**} = \mathbf
{x} + a_{2} \mathbf{g}, a_{2} > a_{1}$ will also place $i$ and $j$ in
different sets.
\end{thmm}

\begin{pf}
Since the final partition at the point~$\mathbf{x}$ placed \textit{i} and
\textit{j} in the same set the partitioning process stopped before
\textit{i} and \textit{j} were separated. Consider any stage and
suppose $B^{*}$ is the set in the partition containing both \textit{i}
and \textit{j}. By assumptions~(i)--(iii), for any $B$ in that
partition we must have $H(A,B\setminus
A;\break\mathbf{x}\,{+}\,a_{1}\mathbf{g})\,{=}\,
H(A,B\setminus A;\mathbf{x})$ \textit{unless} $B \,{=}\, B^{*}$ and $\{i\}
\,{\in}\, A,\break \{j\} \,{\in}\, B^{*}\setminus A$ or $\{j\} \,{\in}\, A, \{i\}
\,{\in}\,
B^{*}\setminus A$. Since $i$ and~$j$ are separated in the final
partition at the point $\mathbf{x} + a_{1} \mathbf{g}$ we must have, at
some stage, $H(A,B^{*}\setminus A;\allowbreak\mathbf{x}+a_{1}\mathbf{g}) >
H(A,B^{*}\setminus A;\allowbreak\mathbf{x})$ for some $A$. It now follows from (i)
that $H(A,B^{*}\setminus A;\mathbf{x}+a_{2}\mathbf{g}) >
H(A,B^{*}\setminus A;\allowbreak\mathbf{x}+a_{1} \mathbf{g})$ for this $A$. Hence
\textit{i} and~\textit{j} will be separated at the point $\mathbf{x} +
a_{2} \mathbf{g}$ at least as early as they were at $\mathbf{x} + a_{1}
\mathbf{g}$.
\end{pf}

We conclude this section with some examples of the partitioning process
using simple models.

\begin{example}[(Treatments versus control in a normal model)]\label{ex4.1}
Let $X_{i}
\sim N(\mu_{i}, 1), i = 1, 2, 3, 4$ be independent. Let
$i = 4$ represent the control population and $i = 1, 2, 3$ represent
the treatment populations. The objective is to test $H_{i4} \dvtx \mu
_i= \mu
_4$ versus $K_{i4} \dvtx \mu_i\neq\mu_4, i = 1, 2, 3$.

To determine an RSD procedure we have opted to begin by taking $ \Omega
$ to be the collection of all sets containing the integer 4 (control)
and at least one other integer chosen from $ \{1, 2, 3\}$. $ \Omega
_{1}$ is the collection of sets containing exactly one integer from
among 1, 2 and 3. $ \Omega_{2}$ is the collection of sets containing
the integer {4}.\vadjust{\goodbreak} As our $H(A, B\setminus A; \mathbf{X})$
function we will use
%
\begin{eqnarray}\label{eq4.2}\qquad
&& H(A, B\setminus A; \mathbf{X})
\nonumber
\\[-8pt]
\\[-8pt]
\nonumber
&&\quad = \biggl|\sum_{j\in A}X_{j}/n(A) - \sum_{j\in
B\setminus A}X_{j}/n(B\setminus A)\biggr| \big/\tau,
\end{eqnarray}
where $\tau= \sqrt{1/n(A) + 1/n(B\setminus A)}$.
\end{example}

We take our three
constants from the \citet{r2} critical values by using
the normal distribution with $\alpha= 0.05$. That is, $C_{1} = 1.48$,
$C_{2} = 1.97$ and  $C_{3} = 2.40 $. To fix ideas we will
take some simple numbers and let $X_{1} = 1 , X_{2} = 4 , X_{3} =
-2$, \mbox{$X_{4} = 0$}.

By our choice of $ \Omega_{1}$ one set must contain
only one integer
and be of the form $A = \{i\}$. Thus at step~1, the RSD procedure
considers the following three possible partitions of $S$:
\begin{longlist}[(iii)]
\item[(i)] $A = \{1\}, S\setminus A = \{2,3,4\},$

\item[(ii)] $A = \{2\}, S\setminus A = \{1,3,4\},$

\item[(iii)] $A = \{3\}, S\setminus A = \{1,2,4\}.$
\end{longlist}

Thus we have $n(A) = 1$
and $n(S\setminus A) = 3$ in all three cases.
When $A = \{i\}$ the function $H$ becomes
\[
H(A, S\setminus A; \mathbf{X}) = \biggl|X_{i} - \sum_{j \neq i} X_{j}/3\biggr|
\Big/\sqrt{4/3}.
\]
In case (i)
\begin{eqnarray*}
H &=& H(\{1\}, \{2,3,4\}; \mathbf{X}) \\
&=& |1 - (4-2+0)/3|/\sqrt{4/3} = 0.29.
\end{eqnarray*}
In case (ii)
\begin{eqnarray*}
H &=& H(\{2\}, \{1,3,4\}; \mathbf{X})\\
& =& |4 - (1-2+0)/3|/\sqrt{4/3} = 3.75.
\end{eqnarray*}
In case (iii)
\begin{eqnarray*}
H& =& H(\{3\}, \{1,2,4\}; \mathbf{X})\\
& = &|-2 - (1+4+0)/3|/\sqrt{4/3} = 3.18.
\end{eqnarray*}

The largest of these is 3.75 which is greater than $2.40 = C_{3}$.
Thus, at step 1, $S$ is split into \{2\} and $\{1, 3, 4\}$ and we continue
to step 2. Next we consider splitting $B = \{1,3,4\}$ into two parts
where the possibilities are

\begin{longlist}[(iv)]
\item[(iv)] $A = \{1\}$, {and} $B\setminus A =
\{3,4\},$

\item[(v)] $A = \{3\}$, {and} $B\setminus A = \{1,4\}.$
\end{longlist}

Thus we have $n(A)
= 1$ and $n(B\setminus A) = 2$ in both cases. When
$A = \{i\}$ the function $H$ becomes
\[
H(A, B\setminus A; \mathbf{X}) = \biggl|X_{i} - \sum_{j \neq i} X_{j}/2\biggr|
\Big/\sqrt{3/2}.\vadjust{\goodbreak}
\]
In case (iv)
\begin{eqnarray*}
H &=& H(\{1\}, \{3,4\}; \mathbf{X}) \\
&=& |1 - (-2+0)/2|/\sqrt{3/2} = 1.63.
\end{eqnarray*}
In case (v)
\begin{eqnarray*}
H &=& H(\{3\}, \{1,4\}; \mathbf{X})\\
 &=& |-2 - (1+0)/2|/\sqrt{3/2} = 2.04.
\end{eqnarray*}

The largest of these is 2.04 which is greater than $1.97 = C_{2}$.
Thus, at step 2, $\{1,3,4\}$ is split into \{3\} and $\{1,4\}$ and we
continue to step 3. At step 3 we consider splitting \{1,4\} into two
parts. $H$ is now simply
\[
H = H(\{1\}, \{4\}; \mathbf{X}) = |1 - (0)|/\sqrt{2} = 0.71.
\]

Since $0.71 < 1.48 = C_{1}$ the set $\{1,4\}$ remains intact and the
process stops. The final partition is $\{2\}$, $\{3\}$ and $\{1,4\}$.
Recalling that if \textit{i} and \textit{j} are placed in different
sets then $H_{ij}$ will be rejected, we find that $H_{14}$ is accepted,
$H_{24}$ is rejected and $H_{34}$ is rejected.

For each (treatment) $i
= 1, 2, 3$ in this setting the interval
property would pertain to the behavior of the test as $X_{i}$ increased
and $X_{k}$ decreased while the other (independent variables) remained
fixed. Thus the vector $\mathbf{g}$ would have a $-1$ in the fourth
position, a $+1$ in the $i$th position and zeroes elsewhere. It is not
difficult to check that the function $H$ given in~(\ref{eq4.2}) satisfies the
conditions of Theorems 1 and~2.

\begin{example}[(Change point in a normal model)]\label{ex4.2}
Let $X_{i} \sim N(\mu_{i},
1), i = 1, \ldots, 10$, be independent. The
objective is to test $H_{i,i+1} \dvtx \mu_i= \mu_{i+1}$ versus
$K_{i,i+1} \dvtx\break
\mu_i\neq\mu_{i+1}, i = 1, \ldots, 9$.
\end{example}

To determine an RSD procedure we
will begin by taking $ \Omega$ to be
the collection of all sets containing at least two consecutive integers
chosen from $ \{1, \ldots, 10\}$. $ \Omega_{1}$ is the collection of
sets containing consecutive integers chosen from among $1, \ldots, 9$.
$\Omega_{2}$ is the collection of sets containing consecutive integers
chosen from $2,\ldots,10$. As our $H(A, B\setminus A; \mathbf{X})$ function
we will again use the function defined in Equation (\ref{eq4.2}). Now there can
be, at most, nine steps in the partition process. Again we can use nine
constants coming from the \citet{r2} critical values by using the normal
distribution with $\alpha= 0.05$.

At step 1 the possible partitions are
%
\begin{eqnarray}
&&A = \{1,\ldots,i\},\quad S\setminus A = \{i+1,\ldots,10\}\nonumber\\
\eqntext{\mbox{for } i
= 1,\ldots,9.}
\end{eqnarray}
Proceeding as in Example \ref{eq4.2} we use the $H$ function and the constant\vadjust{\goodbreak}
$C_{9}$ to decide if and how to divide~$S$. Suppose it is determined
(based on the data) to split $S$ into the sets $\{1,\ldots,d\}$ and
$\{d+1,\ldots,10\}$ for some $d = 1, \ldots, 9$. If $d = 1$, then at
step 2 only $\{2,\ldots,10\}$ is eligible to be split while if $d = 9$,
only $\{1 ,\ldots, 9\}$ is eligible. However, if $1 < d < 9 $, then both
$\{1,\ldots,d\} \mbox{ and } \{d+1,\ldots,10\}$ must be considered at step
2. At step 2 we consider all divisions of the form
%
\begin{eqnarray}
&&A = \{1,\ldots,i\},\quad B\setminus A = \{i+1,\ldots,d\}\nonumber\\
\eqntext{\mbox{for } i
= 1,\ldots,d-1}
\end{eqnarray}
and
%
\begin{eqnarray}
&&A = \{d+1,\ldots,i\},\quad B\setminus A = \{i+1,\ldots,10\}\nonumber\\
\eqntext{\mbox{for } i =
1,\ldots,9.}
\end{eqnarray}

Now using the $H$ functions and the constant $C_{8}$ we would determine
one which, if any, of the above sets should be split. We continue in
this fashion until either there are no more sets eligible to be split
or none satisfy the criterion to be split. As in Example~\ref{eq4.2}, if \textit
{i} and $i+1$ are placed in different sets of the final partition, then
$H_{i,i+1}$ will be rejected.

\section{Multinomial Models}\label{sec5}

In this section we assume that there are $k$ independent multinomial
populations each with $q$ cells. Let $\pi_i , i=1, \ldots,k$ represent
the $i$th population with cell probabilities $p_{ij}, j=1, \ldots, q$.

The individual testing problems are either $H_{i, j} \dvtx\break \pi_i = \pi_{j}$
versus $K_{i,j} \dvtx\pi_i < \pi_{j}$ or $H_{i, j} \dvtx \pi_i = \pi_{j}$
versus $K_{i,j} \dvtx\pi_i \neq\pi_{j}$ where $i < j$. In this case
$\pi_i
< \pi_{j}$ means population $j$ is stochastically larger than
population~$i$, that is,
$ \sum_{l=1}^{m} p_{il}\geq\sum_{l=1}^{m}
p_{jl}$ for $m=1 , \ldots, q$ with some strict inequality.

Let $\widehat{T} (\mathbf{x}_i, \mathbf{x}_j)$ be the two-sample test
statistics used to test $H_{ij}$ that are to be used in the usual
step-down multiple testing procedure. A variety of such test statistics
have been recommended. See, for example, \citet{r1} (BPSS). Most such
statistics, when used to test $H_{i j}$, not as part of a step-down
multiple testing procedure, have the interval property described below.

In this setting it is natural to consider a test's behavior as $x_{i1}$
and $x_{jq}$ both increase while~$x_{iq}$ and~$x_{j1}$ both decrease.
Such changes in data would suggest to a practitioner an ever-increasing
amount of sto\-chastic order. To be precise, suppose $(\mathbf
{x}_i,\mathbf{x}_{j})$ is a reject sample point by virtue of using the
two-sample test $\widehat{\varphi}$. Next, for $ a>0$, consider any
sample point $\mathbf{x}^\ast$ where $x_{\alpha,\beta}^\ast= x_{\alpha
,\beta} + a$ for $(\alpha,\beta) = (i,1)$ and $(\alpha,\beta) = (j,q)$,
$x_{\alpha,\beta}^\ast= x_{\alpha,\beta} - a$ for $(\alpha,\beta) =
(j,1)$ and $(\alpha,\beta) = (i,q)$ and $x_{\alpha,\beta}^\ast=
x_{\alpha,\beta}$ otherwise.\vadjust{\goodbreak} Then $\widehat{\varphi}$ has the interval
property if $\widehat{\varphi}$ also rejects at $(\mathbf{x}_i^\ast,
\mathbf{x}_{j}^\ast)$. In other words $(\mathbf{x}_i^\ast, \mathbf
{x}_{j}^\ast)$ is more indicative of stochastic order than $(\mathbf
{x}_i, \mathbf{x}_{j})$. So if $(\mathbf{x}_i, \mathbf{x}_{j})$ is a
reject point, $(\mathbf{x}_i^\ast, \mathbf{x}_{j}^\ast)$ should also be
a reject point.

Here $\widehat{\varphi}$ has the interval property relative to the
$2q\times1$ vector $\widehat{\mathbf{g}}$ with 1 in positions 1 and
2$q$, $-1$ in positions $q$ and $q+1$ and 0 elsewhere. Thus for the
multiple testing problem the $kq \times1$ vector $\mathbf{g}_{ij}$ has
the value $+1$ in positions $(i-1)(q)+1$ and $(j)(q)$, the value $-1$ in
positions $(i)(q)$ and $(j-1)(q)+1$ and the value 0 in all other positions.

It can be verified that all linear statistics and most nonlinear
statistics listed in BPSS (2009), Section~2.2 have this interval
property. However, these same statistics used as part of a step-down
multiple testing procedure will often lead to induced tests that fail
to have the interval property.

\subsection{Change Point}\label{sec5.1}
In the one-sided change point problem the hypotheses are $H_{i, i+1}
\dvtx
\pi_i\,{=}\, \pi_{i+1}$ versus $K_{i,i+1} \dvtx\pi_i\,{<}\,\pi_{i+1}$, $i=1,
\ldots,
k-1$. That is, in the above $j = i + 1$.
At this point we will demonstrate a simple search that would often lead
to the result that the usual step-down procedure for testing $H_{12}$,
for example, will not have the interval property. That is, if $\varphi
_{12}$ denotes the induced test of $H_{12}$ for the usual step-down
procedure, $\varphi_{12}$ will not have the interval property relative
to $\mathbf{g}_{12}$. The only impediment to this type of search is the
fact that the data consists of integers in each cell and if sample sizes
are small this could be problematic.
An example will follow the recipe.

We follow the pattern exhibited in Figure \ref{fig1} while allowing for the
presence of additional hypotheses (i.e., $k$ can be greater than 3).
Recall that $T_{i(i+1)}(\mathbf{x})$ depends only on $(\mathbf
{x}_i,\mathbf{x}_{i+1})$. Begin by choosing a sample point $\mathbf{x}
= (\mathbf{x}_1^\prime, \mathbf{x}_2^\prime, \ldots, \mathbf{x}_k^\prime
)^\prime$ so that $T_{i(i+1)}(\mathbf{x}) > C_i, i=3, \ldots, k-1;
T_{12}(\mathbf{x})= C_1 + \varepsilon_1, T_{23}(\mathbf{x}) = C_2 +
\varepsilon_2, \varepsilon_1 > 0, \varepsilon_2 >0.$ At $\mathbf{x}$, all
hypotheses are rejected by step-down. Next consider points $\mathbf{x}^{*}$
of the form $\mathbf{x}^{*} = \mathbf{x} + a \mathbf{g}$. That is,
$\mathbf
{x}^\ast= (\mathbf{x}_{1}^{\ast\prime}, \ldots, \mathbf{x}_{k}^{\ast
\prime})^\prime$ where $\mathbf{x}_{i}^\ast= \mathbf{x}_{i}$ for $i=3,
\ldots,k$ but $x_{11}^\ast= x_{11} + a$, $x_{1j}^\ast= x_{1j}$, $j=2, \ldots
, q-1$, $x_{1q}^\ast= x_{1q}-a$, $x_{21}^\ast= x_{21} -
a$, $x_{2j}^\ast= x_{2j}$, $j=2, \ldots, q-1$, $x_{2q}^\ast= x_{2q}+a$.

We note that for most of the statistics used in BPSS (2009) $T_{12}$ is
an increasing function of $a$, $T_{23}$ is a decreasing function of $a$
and $T_{i, i+1}$ for $i\geq3$ does not change with $a$. Choose $a>0$ so
that $T_{23} (\mathbf{x}^\ast) \leq C_2$ and $C_1 + \varepsilon_1 <
T_{12}(\mathbf{x}^\ast) < C_2$. Hence at $\mathbf{x}^{\ast}$ the
step-down procedure would reject $H_{i, i+1}$ for $i\geq3$, but
$H_{12}$ and $H_{23}$ would be accepted. Thus the usual step-down
procedure does not have the interval property in this case.

\begin{example}\label{ex5.1} Consider three independent multinomial distributions, each
with three cells. Test $H_{12}\dvtx\allowbreak  \pi_1 = \pi_2$ versus
$K_{12}\dvtx \pi_1 <
\pi_2$ and $H_{23}\dvtx \pi_2 = \pi_3$ versus $K_{23}\dvtx \pi_2 < \pi
_3$. Use
Wilcoxon--Mann--Whitney (WMW) test statistics $W_{i (i+1)}$ using
midranks. See BPSS (2009). The statistics are then normalized by
letting $Z_{i (i+1)}= [ W_{i (i+1)} - m(m+n +1)/2] /  \sqrt
{mn(m+n+1)/12}$,\break where $m$ and $n$ are the row totals of a two-row table.

For the usual step-down procedure choose constants $C_1= 1.645$ and
$C_{2} = 1.96$. The data in Table~\ref{tab1} offers sample point $\mathbf{x}$.

The statistics are $Z_{12} (\mathbf{x}) = 1.653$ and $Z_{23} (\mathbf
{x}) = 2.006$ leading to rejection of $H_{23}$ followed by rejection of
$H_{12}$. Now we simply choose $a = 1$ to get the sample point $\mathbf
{x}^\ast$ corresponding to Table \ref{tab2}. For~$\mathbf{x}^\ast$,
$Z_{12}(\mathbf{x}^\ast) = 1.954$ and $Z_{23} (\mathbf{x}^\ast) =
1.865.$ The usual step-down procedure now accepts both hypotheses at
$\mathbf{x}^\ast$. Thus the usual step-down procedure with induced test
$\varphi_{12}$ for $H_{12}$ does not have the interval property
relative to $\mathbf{g}_{12}$ where $\widehat{\mathbf{g}}$ has a 1 in
positions 1 and 6, a $-1$ in positions 3 and 4 and 0 elsewhere.
\end{example}

Next we introduce another procedure based on the RSD method that does
have the interval property. Informally, the RSD approach will, at each
stage, consider collections of $2 \times q$ tables formed by collapsing
sets of consecutive rows. It will then apply a two-sample test having
the interval property to these adaptively formed $2 \times q$ tables.
In order to make this precise we need only define the function~$H$ and
the sets $\Omega, \Omega_{1}$ and $\Omega_{2}$. First we take $\Omega$
to be the collection of sets containing at least two consecutive
integers and take $ \Omega_{1} = \Omega_{2}$ to be the collection of
all sets of consecutive integers chosen from $S = \{1, 2, \ldots, k\}
$. Then for any $\widehat{T}$ having the interval property relative to
$\widehat{\mathbf{g}}$ let
\[
H(A, B\setminus A; \mathbf{x}) = \widehat{T}\bigl(Y(A),Y(B\setminus
A)\bigr),
\]
where $Y$ is as defined in Equation (\ref{eq4.1}).

Now we use the current choice of $\mathbf{g}$ along with the
definitions of
$Y$ and $H$ as well as the fact that $\widehat{\mathbf{T}}$ has the
interval property relative to $\widehat{\mathbf{g}}$. This allows us to
verify that assumptions (i)--(iii) of Theorem \ref{thm4.1} are satisfied. Thus
we have

\begin{thmm}\label{thm5.1} RSD has the interval property.
\end{thmm}

To demonstrate the use of the RSD methodology here we apply it to the
model of Example \ref{ex5.1}.

\renewcommand{\theexample}{\arabic{section}.\arabic{example}}
\setcounter{example}{0}
\begin{example}[(Continued)] RSD for the data in Table \ref{tab1}, which
represents sample point $\mathbf{x}$, is carried out as follows: First
Tables \ref{tab3} and \ref{tab4} are formed from Table~\ref{tab1} by averaging frequencies
in rows 1 and 2 for Table \ref{tab3} and averaging rows 2 and 3 for Table \ref{tab4}.

\begin{table}
\caption{Data of Table \protect\ref{tab1} with first two rows combined}\label{tab3}
\begin{tabular*}{\columnwidth}{@{\extracolsep{\fill}}ld{1.1}cd{2.1}c@{}}
\hline
& \multicolumn{1}{c}{\textbf{Same}} &\textbf{Improved}& \multicolumn
{1}{c}{\textbf{Cured}} & \\
\hline
(Placebo${} + {}$Dose 1)${}/{}$2&9.5&226&9.5&245\\
Dose 2&6&196&43&245 \\
\hline
\end{tabular*}
\vspace*{-10pt}
\end{table}

\begin{table}
\caption{Data of Table \protect\ref{tab1} with second two rows combined}\label{tab4}
\begin{tabular*}{\columnwidth}{@{\extracolsep{\fill}}lcccc@{}}
\hline
& \textbf{Same} &\textbf{Improved}& \textbf{Cured} & \\
\hline
Placebo&15&226&\phantom{0}4&245\\
(Dose 1${} + {}$Dose 2)${}/{}$2&\phantom{0}5&211&29&245\\
\hline
\end{tabular*}
\vspace*{-3pt}

\end{table}

At step 1, WMW test statistics $W_{12,3}(\mathbf{x})$ and\break
$W_{1,23}(\mathbf{x})$ are calculated using midranks and then converted
to normalized statistics $Z_{12,3}(\mathbf{x})$ and\break $Z_{1,23}(\mathbf
{x})$. We calculate $Z_{12,3}(\mathbf{x}) = 2.78$ and $Z_{1,23}(\mathbf
{x}) = 2.603$. Using critical values $C_{1} = 1.645$ and $C_{2} = 1.96$
we reject~$H_{23}$ at step 1 based on $Z_{12,3}(\mathbf{x})$. At step 2
we test~$H_{12}$ by using $W_{12}(\mathbf{x})$ normalized to
$Z_{12}(\mathbf{x}) = 1.653$ and thereby reject $H_{12}$ as well. The
sample point $\mathbf{x}^{*}$ is represented by the data in Table~\ref{tab2}.
Proceeding as above we calculate $Z_{12,3}(\mathbf{x}^{*}) = 2.78$ and
$Z_{1,23}(\mathbf{x}^{*}) = 2.824$. This leads to rejection of
$H_{23}$. Next calculate $Z_{12}(\mathbf{x}^{*}) = 1.946$ which leads
to rejection of $H_{12}$.
\end{example}

\subsection{Treatments versus Control}\label{sec5.2}

Let $\pi_k$ be the control population. The hypotheses are $H_{ik}\dvtx
\pi
_i= \pi_k$ versus $K_{ik}\dvtx \pi_i \neq\pi_k, i=1, \ldots, k-1$. Let
$T (\mathbf{x}_i, \mathbf{x}_k)$ be the two-sample test statistics used
for testing $H_{ik}$ that are to be used in the usual step-down testing
procedure. A wide variety of such tests are listed in BPSS (2009). When
we focus on just one hypothesis testing problem we are again comparing
just two populations. Therefore the natural $\widehat{\mathbf{g}}$ is
the same as that defined in the beginning of this section. That is, the
two-sample interval property is relative to the $2q\times1$ vector
$\widehat{\mathbf{g}}$ with 1 in positions 1 and $2q$, $-1$ in
positions $q$ and $q+1$ and 0 elsewhere. For the multiple testing
problem the $kq \times1$ vector $\mathbf{g}_{ik}$ has the value $+1$ in
positions $(i-1)(q)+1$ and $(k)(q)$, the value $-1$ in positions
$(i)(q)$ and $(k-1)(q)+1$ and the value 0 in all other positions.

To show that the usual step-down procedure does not have the interval
property\vadjust{\goodbreak} we follow the pattern exhibited in Figure \ref{fig2} while allowing for
the presence of additional hypotheses (i.e., $k$ can be greater than~3).
Again the discreteness could create a problem with small sample
sizes. Recall that $T_{ik}(\mathbf{x})$ depends only on $(\mathbf
{x}_i,\mathbf{x}_{k})$.

Choose a sample point $\mathbf{x}$ so that $\mathbf{x}_1$ and $\mathbf
{x}_{k}$ are the same, $\mathbf{x}_i, i=3, \ldots, k-1$ are such that
$T_{ik}(\mathbf{x})$ exceeds~$C_i$ by a substantial amount, $\mathbf
{x}_{2}$ is such that $T_{2k}(\mathbf{x}) > C_1 + C_2.$ Thus at $\mathbf
{x}, H_{2k}$ is accepted. Now choose $\mathbf{x}^{*}$ so that $C_1 <
T_{1k} (\mathbf{x}^\ast) < C_2$, and $T_{2k}(\mathbf{x}^\ast) = C_2 +
\varepsilon$. This is possible since $T_{1k}$ has the interval property
and since $\mathbf{x}_2^\ast$ is closer to $\mathbf{x}_k^\ast$ than
$\mathbf{x}_2$ is to~$\mathbf{x}_k$. Now at $\mathbf{x}^\ast$ the
procedure rejects $H_{1k}$ and~$H_{2k}$. \mbox{Finally} choose $\mathbf
{x}^{\ast\ast}$ so that $T_{2k} (\mathbf{x}^{\ast\ast}) \leq C_2$
and\break
$T_{1k}(\mathbf{x}^{\ast\ast}) \leq C_2$. This is possible since
$\mathbf{x}^{\ast\ast}$ is such that~$\mathbf{x}_1^{\ast\ast}$ and~$\mathbf{x}_k^{\ast\ast}$ are moving further apart while~$\mathbf
{x}_2^{\ast\ast}$ and $\mathbf{x}_k^{\ast\ast}$ are moving closer to
each other. Thus at~$\mathbf{x}_2^{\ast\ast}$, $H_{1k}$ and $H_{2k}$ are
accepted. This demonstrates that the usual step-down procedure lacks
the interval\break property relative to $\mathbf{g}$.

Now we indicate the RSD method that does have the interval property.
Informally, the RSD approach will, at each stage, consider collections
of $2 \times q$ tables formed by taking one row to be one of the
treatments while the other row is the result of combining all other
treatments with the control. It will then apply a two-sample test
having the interval property to these adaptively formed $2 \times q$
tables. In order to make this precise we need only define the function~$H$ and the sets $\Omega, \Omega_{1}$ and $\Omega_{2}$. First we take $
\Omega$ to be the collection of all sets containing $k$ and at least
one other integer chosen from $ \{1, 2, \ldots, k-1\}$. $ \Omega_{1}$
is the collection of sets containing exactly one integer. $ \Omega_{2}$
is the collection of sets containing the integer $k$. Then for any
$\widehat{T}$ having the interval property relative to $\widehat{\mathbf
{g}}$ let
\[
H(A, B\setminus A; \mathbf{x}) = \widehat{T}\bigl(Y(A),Y(B\setminus A)\bigr).
\]

Now we use the current choice of $\mathbf{g}$ along with the
definitions of
$Y$ and $H$ as well as the fact that $\mathbf{T}$ has the interval property
relative to $\widehat{\mathbf{g}}$. This allows us to verify that
assumptions (i)--(iii) of Theorem \ref{thm4.2} are satisfied. Thus we have

\begin{thmm}\label{thm5.2}
RSD has the interval property.
\end{thmm}

\subsection{All Pairwise Differences}\label{sec5.3}
The hypotheses are $H_{ij}\dvtx \pi_i= \pi_j$ versus $K_{ij}\dvtx \pi
_i\neq\pi
_j, i=1, \ldots, k-1, j=i+1, \ldots, k.$ Once again it can be
shown that the usual step-down procedure does not have the interval
property in this case. Focusing on\vadjust{\goodbreak} $H_{12}$ and utilizing statistics
$T_{12} \mbox{ and } T_{23}$ as in the arguments of Section \ref{sec5.1} will
suffice to give the results in this case.

We now offer an RSD procedure that does have the interval property. The
basis of this RSD procedure is the PADD procedure for testing all
pairwise normal means in CSC (2010). For the multinomial case we
describe the procedure now.

Again it suffices to follow the exposition in Section~\ref{sec3}. Here we let
$\Omega$ be the collection of all sets containing at least two
integers. Further let $\Omega_{1} = \Omega_{2}$ be the collection of
all nonempty subsets of $S = \{1, 2, \ldots, k\}$. Next take
\[
H(A,B\setminus A;\mathbf{x}) = \widehat{T}\bigl(Y(A;\mathbf{x}),Y(B\setminus
A;\mathbf{x})\bigr),
\]
where $\widehat{T}$ is any test statistic for testing independence in a
$2\times q$ table that has the interval property relative to $\widehat
{\mathbf{g}}$.

The interpretation is as follows: By definition every $Y(A;\mathbf{x})$
will be the result of combining all rows corresponding to indices in
$A$. In determining how a~set $B$ might be split we look at every
possible way to collapse all the rows corresponding to the indices in
$B$ into just two rows. Then a test is performed for each resulting
$2\times q$ table. For example, if $k = 4$ and $B = \{1, 2, 3, 4\}$,
then the possible splits are $\{1\}$ {and} $\{2, 3, 4\}$, $\{2\}$
{and} $\{1, 3, 4\}$, $\{3\}$ {and} $\{1, 2, 4\}$, $\{4\}$ {and} $\{
1, 2, 3\} $,
$\{1, 2\}$ {and} $\{3, 4\}$, $\{1, 3\}$ {and} $\{2, 4\}$ {or} $\{1, 4\}$ {and} $\{2, 3\}$.

With these definitions one can check that assumptions (i)--(iii) of
Theorem \ref{thm4.2} are satisfied. Thus we have

\begin{thmm}\label{thm5.3}
RSD has the interval property.
\end{thmm}

\section{Multivariate Normal Models}\label{sec6}

Let $\mathbf{x}_{i}, i = 1, \ldots,k$, be independent \textit
{q}-variate normal random vectors with mean vectors $\bolds{\mu}
_{i}$ and known nonsingular covariance matrix $\Sigma$. All hypotheses
are concerned with pairwise differences between mean vectors. In light
of this we assume without loss of generality that $\Sigma= \mathrm{I}$.
The two-sample test statistic that will serve as the basis for all
usual step-down procedures considered to test $H_{ij}\dvtx \bolds
{\mu}
_{i} = \bolds{\mu}_{j}$ versus $K_{ij}\dvtx \bolds{\mu}_{i}
\neq
\bolds{\mu}_{j}$ is
%
%
\begin{equation}\label{eq6.1}
\widehat{T}(\mathbf{x}_{i}, \mathbf{x}_{j}) = (\mathbf{x}_{i} - \mathbf{x}_{j})'(\mathbf{x}_{i} - \mathbf{x}_{j})/2
\end{equation}
which has a chi-squared distribution with \textit{q} degrees of freedom.

Here a natural form of the interval property is along points
%
\begin{eqnarray}\label{eq6.2}
\mathbf{x} &=& (\mathbf{x}_1^\prime, \mathbf{x}_2 ^\prime, \ldots,
\mathbf{x}_k ^\prime)^\prime,
\end{eqnarray}
\begin{eqnarray}
\label{eq6.3}\mathbf{x}^\ast&=& \bigl((\mathbf{x}_{1}-r_{1} \mathbf{1})^\prime, (\mathbf
{x}_{2}+ r_{1}\mathbf{1})^\prime, \mathbf{x}_3^\prime, \ldots,\mathbf
{x}_k^\prime\bigr)^\prime,
\\
\label{eq6.4}\qquad\mathbf{x}^{\ast\ast} &=& \bigl((\mathbf{x}_{1}-r_{2} \mathbf{1})^\prime,
(\mathbf{x}_{2}+ r_{2}\mathbf{1})^\prime, \mathbf{x}_3^\prime, \ldots
,\mathbf{x}_k^\prime\bigr)^\prime,
\end{eqnarray}
where $0< r_1 < r_2$ and $\mathbf{1}$ is a vector of all 1's. Thus
$\widehat{\mathbf{g}} = (-1,\ldots, -1, 1, \ldots, 1)^\prime$ and
$\mathbf
{g}$ has entries of $-1$ for coordinates corresponding to population
\textit{i}, 1 for coordinates corresponding to population \textit{j}
and 0 elsewhere.\vspace*{2pt}

\subsection{All Pairwise Differences}\label{sec6.1}\vspace*{2pt}

The case of $q=1$ has been studied by CSC (2010). For arbitrary \textit
{q}, the lack of the interval property of the usual step-down procedure
is shown by focusing on $H_{12}$ and utilizing statistics $T_{12},
T_{23}$ as in the argument of Section \ref{sec5.1}.

At this point we describe an RSD which does have the interval property.
Here we let $\Omega$ be the collection of all sets containing at least
two integers. Further let $\Omega_{1} = \Omega_{2}$ be the collection
of all nonempty subsets of $S = \{1, 2, \ldots, k\}$. Next take
\begin{eqnarray*}
&&H(A,B\setminus A;\mathbf{x}) \\
&&\quad= \widehat{T}\bigl(Y(A;\mathbf{x})/n(A)
,Y(B\setminus A;\mathbf{x})/n(B\setminus A)\bigr)\\
&&\qquad{}/\bigl(1/n(A)+1/n(B\setminus A)\bigr).
\end{eqnarray*}

Again the assumptions of Theorem \ref{thm4.2} can be verified and the interval
property established.\vspace*{2pt}

\subsection{Change Point}\label{sec6.2}\vspace*{2pt}

The hypotheses are $H_{i(i+1)}\dvtx \bolds{\mu}_i = \bolds
{\mu}_{i+1}$ versus\break $K_{i(i+1)}\dvtx \bolds{\mu}_i \neq\bolds{\mu}_{i+1},
i=1,2, \ldots, k-1$. Test statistics for the usual step-down procedure
are $\widehat{T}(\mathbf{x}_{i}, \mathbf{x}_{i+1})$ as given in (\ref{eq6.1}).
The lack of the interval property for the usual step-down is shown by
focusing on $H_{12}$ and utilizing statistics $T_{12}$ and $T_{23}$ as
in the argument of Section \ref{sec5.1}. Here again we let $\mathbf{x}, \mathbf
{x}^\ast, \mathbf{x}^{\ast\ast}$ be as in (\ref{eq6.2}), (\ref{eq6.3}) and (\ref{eq6.4}).

For RSD we proceed as follows: Take $\Omega$ to be the collection of
sets containing at least two consecutive integers and take $ \Omega_{1}
= \Omega_{2}$ to be the collection of all sets of consecutive integers
chosen from $S = \{1, 2, \ldots, k\}$ and again choose
\begin{eqnarray*}
&&H(A,B\setminus A;\mathbf{x})\\
&&\quad= \widehat{T}\bigl(Y(A;\mathbf{x})/n(A)
,Y(B\setminus A;\mathbf{x})/n(B\setminus A)\bigr)\\
&&\qquad{}/\bigl(1/n(A)+1/n(B\setminus A)\bigr).
\end{eqnarray*}

\begin{table*}[b]
\vspace*{-6pt}
\tabcolsep=0pt
\caption{Performance of RSD and SU. The mean of the control population
is 0.0. Each mean value listed represents five treatments. All
unspecified means are equal to 0.0}\label{tab5}
\begin{tabular*}{\textwidth}{@{\extracolsep{4in minus 4in}}lcd{2.2}ccd{2.1}d{2.1}d{2.1}d{2.1}cc@{}}
\hline
&&&\multicolumn{6}{@{}c@{}}{\textbf{Expected number of errors}} &&\\
\ccline{4-9}\\[-6pt]
\multicolumn{3}{@{}l@{}}{\textbf{Means for treatment number}} &\multicolumn
{2}{c}{\textbf{Type I}}&\multicolumn
{2}{c}{\textbf{Type II}}&\multicolumn{2}{c}{\textbf{Total}}&\multicolumn
{2}{c@{}}{\textbf{FDR}}\\
\ccline{1-3,4-5,6-7,8-9,10-11}\\[-6pt]
\multicolumn{1}{@{}l}{\textbf{1--5}}&
\multicolumn{1}{c}{\textbf{6--10}}& \multicolumn{1}{c}{\textbf{11--15}}
& \multicolumn{1}{c}{\textbf{RSD}}&
\multicolumn{1}{c}{\textbf{SU}}& \multicolumn{1}{c}{\textbf{RSD}} &
\multicolumn{1}{c}{\textbf{SU}}&
\multicolumn{1}{c}{\textbf{RSD}}&
\multicolumn{1}{c}{\textbf{SU}}&\multicolumn{1}{c}{\textbf
{RSD}}&\multicolumn{1}{c@{}}{\textbf{SU}} \\
\hline
0.00 & 0.00 & 0.00&0.1&0.7& 0.0& 0.0& 0.1& 0.7&
0.048& 0.045\\
0.00 & 0.00 & -2.00&0.1&0.7& 3.5& 4.4& 3.6& 5.1&
0.046& 0.050\\
0.00 & 0.00 & -4.00&0.3&0.8& 0.0& 0.8& 0.4& 1.6&
0.051& 0.054\\
0.00 & 2.00 & -2.00&0.3&0.7& 6.0& 8.8& 6.2& 9.5&
0.045& 0.044\\
0.00 & 2.00 & 2.00&0.2&0.8& 6.8& 8.5& 7.0& 9.2&
0.048& 0.044\\
0.00 & 2.00 & -4.00&0.4&1.0& 2.7& 4.6& 3.1& 5.6&
0.049& 0.054\\
0.00 & 2.00 & 4.00&0.4&0.8& 2.7& 4.8& 3.2& 5.6&
0.048& 0.048\\
0.00 & 4.00 & -4.00&0.6&0.9& 0.0& 1.0& 0.6& 1.9&
0.050& 0.052\\
0.00 &4.00 & 4.00&0.6&0.9& 0.0& 1.1& 0.6& 2.0&
0.049& 0.050\\
2.00 & 2.00 & -2.00&0.4&0.9& 8.1&12.8& 8.5& 13.7& 0.045&
0.048\\
2.00 & 2.00 & 2.00&0.4&0.9& 10.0&12.3& 10.3& 13.2&
0.055& 0.045\\
2.00 & 2.00 & -4.00&0.6&0.9& 5.3& 8.2& 5.9& 9.2&
0.051& 0.048\\
2.00 &2.00 & 4.00&0.6&0.9& 5.3& 8.6& 5.9& 9.4&
0.034& 0.047\\
2.00 & 4.00 & -4.00&0.7&1.1& 2.3& 4.6& 3.0& 5.7&
0.049& 0.052\\
2.00 & 4.00 &4.00&0.7&1.0& 2.3& 4.7& 3.0& 5.7&
0.049& 0.049\\
4.00 & 4.00 & -4.00&0.8&1.2& 0.0& 1.1& 0.8& 2.3&
0.048& 0.050\\
4.00 & 4.00 & 4.00&0.8&1.3& 0.0& 1.3& 0.9& 2.6&
0.050& 0.055\\
\hline
\end{tabular*}
\end{table*}

Once again the assumptions of Theorem \ref{thm4.2} can be verified and so RSD
has the interval property in this case.

\begin{remark}\label{rem6.1} For the univariate normal change point problem, MRD is a
special case of an RSD procedure. For a numerical simulation study
comparing MRD with step-down see \citet{r11}.\vspace*{2pt}
\end{remark}

\subsection{Treatments versus Control}\label{sec6.3}\vspace*{2pt}

The case $q=1$ is treated in CSX (2009)  and the case of arbitrary $q$
was treated in Cohen, Sackrowitz and Xu (\citeyear{r10}) (CSX).

The hypotheses are $H_{ik} \dvtx\bolds{\mu}_i = \bolds{\mu}_k$
versus $K_{ik}\dvtx \bolds{\mu}_i \neq\bolds{\mu}_k$, $i= 1, 2,
\ldots, k-1$. The usual step-down two-sample statistics at step 1 are
$T_{ik} = (\mathbf{x}_i - \mathbf{x}_k)^\prime(\mathbf{x}_i - \mathbf
{x}_k) / 2.$ To determine the RSD procedure we take $ \Omega$ to be
the collection of all sets containing the integer $k$ and at least one
other integer chosen from $ \{1, 2, \ldots, k-1\}$. $ \Omega_{1}$ is
the collection of sets containing exactly one integer from among $\{1,
\ldots, k-1\}$. $ \Omega_{2}$ is the collection of sets containing the
integer \textit{k}. As in Sections~\ref{sec6.1} and~\ref{sec6.2} let
%
\begin{eqnarray}\label{eq6.5}
\qquad && H(A,B\setminus A;\mathbf{x}) \nonumber\\
&&\quad= \widehat{T}\bigl(Y(A;\mathbf{x})/n(A)
,Y(B\setminus A;\mathbf{x})/n(B\setminus A)\bigr)\\
&&\qquad{}/\bigl(1/n(A)+1/n(B\setminus A)\bigr).\nonumber
\end{eqnarray}

The RSD we use in this situation is simply the vector analog to the
procedure shown in Example~\ref{ex4.1}. Now, of\vadjust{\goodbreak} course, $q \geq1$, scalar
variables and parameters become vectors and the number of treatments is
$k-1$. For the function $H$ we use the vector analog to~(\ref{eq4.2}) that is
given in (\ref{eq6.5}). Implementation follows the same steps as in Example
\ref{eq4.1}. The only difference might be in the choice of constants as
discussed below.

Here again it can be shown that the usual step-down test of $H_{ik}$
does not have the interval property when $\mathbf{g} = (\mathbf{0},
\ldots, \mathbf{0}, -\mathbf{1}, \mathbf{0}, \ldots, \mathbf{0},
\mathbf{1})$ with the $-\mathbf{1}$ in the $i$th position while RSD does
have the interval property.

We now give two simple examples of how the RSD method might be
constructed and used.
First we mention that for the standard step-up procedure the \citet{r3}
constants in the two-sided case are given by
%
\begin{equation}\label{eq6.6}
C^{\mathrm{BH}}_{i} = \Phi^{-1}\bigl(1 - (k+1-i)(\alpha/2) /k\bigr).
\end{equation}

The constants given in \citet{r2} are
%
%
\begin{equation}\label{eq6.7}
 C^{\mathrm{BG}}_{i} = \Phi^{-1}\bigl(1 - i(\alpha/2)
 /(k+1-i(1-\alpha/2)\bigr).\hspace*{-25pt}
\end{equation}

\begin{table*}
\vspace*{-6pt}
\tabcolsep=0pt
\caption{Performance of RSD and SU. The mean of the control population
is 0.0. Each mean value listed represents eight treatments. All
unspecified means are equal to 0.0}\label{tab6}
\begin{tabular*}{\textwidth}{@{\extracolsep{4in minus
4in}}lcd{2.2}ccd{2.1}d{2.1}d{2.1}d{2.1}cc@{}}
\hline
&&&\multicolumn{6}{c}{\textbf{Expected number of errors}} &&\\
\ccline{4-9}\\[-6pt]
\multicolumn{3}{@{}l}{\textbf{Means for treatment number}} &\multicolumn
{2}{c}{\textbf{Type I}}&\multicolumn
{2}{c}{\textbf{Type II}}&\multicolumn{2}{c}{\textbf{Total}}&\multicolumn
{2}{c@{}}{\textbf{FDR}}\\
\ccline{1-3,4-5,6-7,8-9,10-11}\\[-6pt]
\multicolumn{1}{@{}l}{\textbf{1--8}}&
\multicolumn{1}{c}{\textbf{9--16}}& \multicolumn{1}{c}{\textbf{17--24}}
& \multicolumn{1}{c}{\textbf{RSD}}&
\multicolumn{1}{c}{\textbf{SU}}& \multicolumn{1}{c}{\textbf{RSD}} &
\multicolumn{1}{c}{\textbf{SU}}&
\multicolumn{1}{c}{\textbf{RSD}}&
\multicolumn{1}{c}{\textbf{SU}}&\multicolumn{1}{c}{\textbf
{RSD}}&\multicolumn{1}{c@{}}{\textbf{SU}} \\
\hline
0.00 & 0.00 & 0.00&0.0&0.5& 0.0& 0.0& 0.0& 0.5&
0.031& 0.038\\
0.00 & 0.00 & -2.00&0.1&0.7& 6.1& 6.9& 6.2& 7.6&
0.029& 0.046\\
0.00 & 0.00 & -4.00&0.3&0.8& 0.0& 0.9& 0.3& 1.8&
0.031& 0.051\\
0.00 & 2.00 & -2.00&0.2&0.8& 9.6&13.7& 9.8& 14.5& 0.027&
0.043\\
0.00 & 2.00 & 2.00&0.2&0.8& 12.1&13.0& 12.3& 13.8&
0.037& 0.043\\
0.00 & 2.00 & -4.00&0.4&1.1& 4.5& 6.7& 4.9& 7.8&
0.030& 0.051\\
0.00 & 2.00 & 4.00&0.4&1.0& 4.6& 6.9& 5.0& 7.9&
0.029& 0.048\\
0.00 & 4.00 & -4.00&0.5&1.4& 0.0& 1.2& 0.6& 2.6&
0.030& 0.056\\
0.00 & 4.00 & 4.00&0.5&1.3& 0.0& 1.3& 0.6& 2.6&
0.030& 0.053\\
2.00 & 2.00 & -2.00&0.3&1.0& 13.3&19.6& 13.6& 20.6& 0.028&
0.045\\
2.00 & 2.00 & 2.00&0.3&1.0& 19.2&18.8& 19.5& 19.7&
0.058& 0.040\\
2.00 & 2.00 & -4.00&0.5&1.0& 9.4&12.1& 9.9& 13.1& 0.034&
0.045\\
2.00 & 2.00 & 4.00&0.5&1.0& 9.4&12.5& 10.0& 13.5&
0.034& 0.045\\
2.00 & 4.00 & -4.00&0.6&1.3& 3.8& 6.5& 4.5& 7.8&
0.030& 0.048\\
2.00 & 4.00 & 4.00&0.6&1.2& 3.8& 6.7& 4.5& 7.9&
0.029& 0.046\\
4.00 & 4.00 & -4.00&0.8&1.4& 0.0& 1.3& 0.8& 2.7&
0.030& 0.047\\
4.00 & 4.00 & 4.00&0.8&1.6& 0.0& 1.4& 0.8& 3.0&
0.030& 0.052\\
\hline
\end{tabular*}
\end{table*}

Take $q = 1$ and $k = 101$ so we have 100 treatments and one
control.
Suppose further that the only reasonable scenario is that the number of
truly significant treatments is sparse, say, at the very most, 15\% of
the treatments. Table \ref{tab5} gives the results of a~simulation\vadjust{\goodbreak} using 5000
iterations at each parameter point. We compare the RSD method with
step-up on the criteria of FDR, the expected number of Type~I errors
and the expected number of Type II errors. For RSD we were able to use
the critical values of~(\ref{eq6.7}) with $\alpha= 0.05$ without any
modification. For step-up, on the other hand, using $\alpha= 0.05$ in
(\ref{eq6.6}) resulted in a procedure that was (due to the dependence) too
conservative and put it at a disadvantage. Instead we found, using
simulation, that taking $\alpha= 0.07$ in (\ref{eq6.6}) gave a better
performing procedure for this covariance structure. For this
application RSD has the interval property, is comparable to step-up
relative to FDR and makes fewer mistakes than step-up. Table \ref{tab6} allows
for a less sparse situation allowing as many as 24\% better treatments.
Here simulation indicated that we should again take $\alpha= 0.07$ in
(\ref{eq6.6}) for step-up and the critical values of RSD should correspond to
$\alpha= 0.03$ in (\ref{eq6.7}).

In both Tables \ref{tab5} and \ref{tab6} the mean of the control population is taken
to be 0.0. In Table \ref{tab5} the means given in the first three columns each
represent five treatment means. The other 85 treatment means are 0.0.
For example, in the next to last row, the first 10 treatment means
would be 4.00 and the next five treatment means would be $-4.00$. In this
case 15\% of the treatments would be nonzero. In Table \ref{tab6} the means
given in the first three columns each represent eight treatment
means.\vadjust{\goodbreak}
Thus the maximum number of nonzero treatment means would be, at most,
24\%. Note both Tables~\ref{tab5} and~\ref{tab6} indicate fewer errors for RSD for
all parameter points considered.

\begin{remark}\label{rem6.2}
For the univariate normal treatments versus control problem
MRD is a special case and natural choice of an RSD procedure. One of
the simulation studies in \citet{r11} was done for this same model but
for many more treatments. Both step-up and step-down were considered.
As described in that paper it was more difficult to arrive at
appropriate choices for critical values. The nature of the results was
the same but, due to the large number of populations, the results were stronger.
\end{remark}

\section{Nonparametric Models}\label{sec7}

Nonparametric multiple testing is discussed in\break \citet{r14}. Here we
begin with~$n$ independent observations from each of $k$ independent
populations $F_1, \ldots, F_k$. The collection of all \textit{nk}
observations are ranked and we let $R_i = $ the average of the ranks
for the observations coming from population $i$. Also let $\mathbf{R} =
(R_1, \ldots, R_k)^\prime$. For testing $H_{ij}\dvtx F_i = F_j$ versus
$K_{ij}\dvtx F_i < F_j$ or $H_{ij}\dvtx F_i = F_j$ versus
$K_{ij}\dvtx F_i \neq
F_j$ based on $\mathbf{R}$ it is natural to study the behavior of testing
procedures as $R_i$ decreases and $R_j$ increases.\vadjust{\goodbreak}

This model fits our original setting with $R_i$ playing the role of
$\mathbf{x}_i$ and $q = 1$. Here $\widehat{\mathbf{g}} = (-1,
1)^\prime$ and $\mathbf{g}$ is the $k \times1$ vector with $-1$ as the
$i$th coordinate, 1 as the $j$th coordinate and 0 elsewhere.

\subsection{All Pairwise Differences}\label{sec7.1}

The problem of nonparametric multiple testing of all pairwise
comparisons of distributions has been treated by \citet{r8} (CS). There
it is shown that the step-down procedure of \citet{r4} based on ranks
lacks an interval property. It is also shown in CS (2010) that the RSD
procedure (called RPADD there) does have the interval property.

\subsection{Change Point}\label{sec7.2}
Next we consider testing $H_{i(i+1)} \dvtx F_i = F_{i+1}$ versus
$K_{i(i+1)}\dvtx F_i < F_{i+1}, i=1, \ldots, k-1$ assuming $F_i \leq
F_2 \leq\cdots\leq F_k$. Assume sample sizes are $n$ for each
population. It is possible to show that a typical step-down procedure
using two-sample rank tests (based on separate ranks or joint ranks)
for $H_{i(i+1)}$ would not have the interval property. However, the RSD
procedure which we now describe will have the interval property. As in
the other change point settings, take $\Omega$ to be the collection of
sets containing at least two consecutive integers and take $ \Omega_{1}
= \Omega_{2}$ to be the collection of all sets of consecutive integers\
chosen from $S = \{1, 2, \ldots, k\}.$ Here we let
\begin{eqnarray*}
&& H(A,B\setminus A;\mathbf{R}) \\
&&\quad= \bigl(Y(A;\mathbf{R})/N(A) \\
&&\hspace*{2pt}\qquad{}- Y(B\setminus
A;\mathbf{R})/N(B\setminus A)\bigr)/\sigma_{A,B},
\end{eqnarray*}

where
\begin{eqnarray*}
\sigma_{A,B}^{2} &= &w\bigl(1/N(A)+1/N(B\setminus A)\bigr)/12 \quad\mbox{and}\\
 w &=& k(kn+1).
\end{eqnarray*}
With these definitions it is easy to verify the conditions of Theorem
\ref{thm4.1} to obtain

\begin{thmm}\label{thm7.1} RSD has the interval property for testing $H_{i,i+1}.$
\end{thmm}

\subsection{Treatments versus Control}\label{sec7.3}

For testing treatments versus control the hypotheses are $H_{ik} \dvtx F_i
= F_k$ versus $K_{ik}\dvtx F_i \neq F_k$. Now consider the usual step-down
procedure which is based on the two-population statistic
\[
T_{ik} = |R_i - R_k| / \sigma_{\{i\},\{k\}}
\]
in comparing the $i$th treatment with the control. It can be shown that
the usual step-down procedure does not have the interval property for
testing $H_{ik}$.\vadjust{\goodbreak}

On the other hand, it can be shown that the RSD procedure for this
model does have the interval property for testing $H_{ik}$. RSD in this
case is defined as follows: Let $ \Omega$ be the collection of all sets
containing $k$ and at least one other integer chosen from $S = \{1, 2,
\ldots, k-1\}$. $ \Omega_{1}$ is the collection of sets containing
exactly one integer. $ \Omega_{2}$ is the collection of sets containing
the integer \textit{k}. Then take
\begin{eqnarray*}
&& H(A,B\setminus A;\mathbf{R}) \\
&&\quad= \bigl|Y\bigl(A;\mathbf{R} /N(A)\bigr) \\
&&\hspace*{2pt}\qquad{}- Y(B\setminus
A;\mathbf{R})/N(B\setminus A)\bigr| /\sigma_{A,B},
\end{eqnarray*}
where $\sigma_{A,B}^{2}$ is as defined in Section \ref{sec7.2} above.
With these definitions it is easy to verify the conditions of Theorem
\ref{thm4.1} to obtain
\begin{thmm}\label{thm7.2} RSD has the interval property for testing $H_{i,k}.$
\end{thmm}

\section*{Acknowledgment}
Research supported by NSF Grant 0894547 and NSA Grant H-98230-10-1-0211.

%

\end{document}